\documentclass[journal]{IEEEtran}

 \usepackage[dvips]{graphicx}
   \usepackage{psfrag}
 \DeclareGraphicsExtensions{.eps}

\usepackage{array}
\usepackage{cite}
\usepackage{amsmath}
\usepackage{amssymb}
\newtheorem{property}{Property}
\usepackage{enumitem}
\usepackage{fancyhdr}
\usepackage[absolute]{textpos}
\newcommand{\copyrightstatement}{
    \begin{textblock}{15}(0.5,0.3)    % tweak here: {box width}(leftposition, rightposition)
         \noindent
         \centering
         \textblockcolour{white}
         \footnotesize
         \copyright 2016 IEEE. Personal use of this material is permitted. Permission from IEEE must be obtained for all other uses, in any current or future media, including reprinting/republishing this material for advertising or promotional purposes, creating new collective works, for resale or redistribution to servers or lists, or reuse of any copyrighted component of this work in other works
    \end{textblock}
}
\usepackage{url}
\begin{document}

\copyrightstatement

\title{Modeling the Energy Consumption \\ of the HEVC Decoding Process}

\author{Christian~Herglotz,
        Dominic~Springer,
        Marc~Reichenbach,
        Benno~Stabernack,
        and~Andr\'e~Kaup,~\IEEEmembership{Fellow,~IEEE}% <-this % stops a space
\thanks{Manuscript received December 8, 2015; revised June 14, 2016, accepted July 11, 2016. This work was financed by the Research Training Group 1773 ``Heterogeneous Image Systems'', funded by the German Research Foundation (DFG).}
\thanks{C. Herglotz, D. Springer, and A. Kaup are with the Chair of Multimedia Communications and Signal Processing, Friedrich-Alexander-Universit\"at Erlangen-N\"urnberg (FAU), Erlangen, Germany (\{ christian.herglotz, dominic.springer, andre.kaup \} @FAU.de). }
\thanks{M. Reichenbach is with the Chair of Computer Science 3, Friedrich-Alexander-Universit\"at Erlangen-N\"urnberg (FAU), Erlangen, Germany (marc.reichenbach@informatik.uni-erlangen.de).}
\thanks{ B. Stabernack is with the Fraunhofer Heinrich-Hertz-Institut (HHI), Berlin, Germany (benno.stabernack@hhi.fraunhofer.de). }% <-this % stops a space
%\thanks{}
}

\maketitle

\begin{abstract} 
In this paper, we present a bit stream feature based energy model that accurately estimates the energy required to decode a given HEVC-coded bit stream. Therefore, we take a model from literature and extend it by explicitly modeling the in-loop filters, which was not done before. Furthermore, to prove its superior estimation performance, it is compared to seven different energy models from literature. By using a unified evaluation framework we show how accurately the required decoding energy for different decoding systems can be approximated. We give thorough explanations on the model parameters and explain how the model variables are derived. To show the modeling capabilities in general, we test the estimation performance for different decoding software and hardware solutions, where we find that the proposed model outperforms the models from literature by reaching frame-wise mean estimation errors of less than $7\%$ for software and less than $15\%$ for hardware based systems. 
\end{abstract}


\thispagestyle{fancy}

\section{Introduction}
\label{sec:intro}
\IEEEPARstart{D}{uring} the last two decades, video coding technologies have gone through a remarkable evolution. Former, rather simplistic approaches aiming at low resolution and low quality video content have evolved into high performance, complex and efficient coding techniques that are capable of reducing bit rates for Ultra High Definition (UHD) resolution videos. In the meantime, the common use-case for video coding switched from traditional desktop personal computer (PC) applications to various different specialized devices such as televisions, cameras, tablet PCs, smartphones, or other embedded devices. These place new constraints on the encoding and decoding solutions and range from low complexity and portability to ease of integration and low energy consumption. 

In this paper, we focus on the decoding energy consumption that is typically important for portable devices such as tablet PCs or smartphones. Nowadays, these devices are able to perform real-time streaming and decoding of video content such that the user can watch videos anywhere and anytime. Due to the high complexity of the decoder, this process is one of the portable device's major power consumers, shortening the operating time until the battery has to be recharged. 
In \cite{Carroll10} it is shown that a smartphone requires about a quarter of its power consumption merely for the decoding part, for a rather outdated H.263 decoder solution (display at medium brightness). Our measurements show that for the state-of-the-art High Efficiency Video Coding (HEVC) codec \cite{Sullivan12}, the real-time software decoding of a High Definition (HD) sequence would drain the smartphone's battery in about four hours, not taking into account peripheral energy consumers such as the display. 

Hence, research aiming at reducing the amount of energy required to decode a video stream is a worthwhile task. Therefore, a common approach is to develop specialized decoding hardware such as the HEVC-decoding chip presented in \cite{Huang13}. % for a preliminary draft of the HEVC standard. 
Furthermore, for the different modules of the decoder, several optimized hardware solutions have been proposed \cite{Zhu13,Kalali12,Jiang14,Palomino12,Park13}. %,Tikekar14,Do14}. 
Other approaches aim at the optimization of existing solutions \cite{vdSchaar05, Lee07}. To this end, one proposal was to determine the decoder's complexity, which can be seen as an approximation to the energy consumption, to then be incorporated into the Rate-Distortion Optimization (RDO) process on the encoder side. In this context, van der Schaar et al. \cite{vdSchaar05} proposed a generic complexity model for typical decoder processes and Lee et al. \cite{Lee07} derived a dedicated model for the H.264 decoder that considers cache misses, interpolation filters, and the number of motion vectors per macroblock. Using this information inside the RDO-process helps the encoder to construct a video bit stream of low decoding complexity. As a drawback, the encoder only has incomplete information about the decoder's behavior when using this model. 

Although complexity can be seen as a close approximation to energy consumption, different solutions have been proposed using the actual power or energy consumption. Li et al. \cite{Li12} proposed a power model based on the high-level bit stream parameters resolution, frames per second, and quantization. In the paper it is shown that the bit stream can be encoded in such a way as to match an energy requirement given by the decoding device by adaptively changing these parameters in the encoder. This approach was expounded upon by Raoufi et al. \cite{Raoufi13} by incorporating the transmission of the bit stream via a wireless channel and concentrating on bit rate and rate of intra coded frames. 

The models presented before were all established for the H.264/AVC (Advanced Video Coding) standard \cite{ITU_H.264}. To estimate the decoding energy of HEVC decoders, Ren et al. \cite{Ren14b} selected the processor-level parameters instructions, cache misses, and hardware interrupts as a foundation and achieved average relative estimation errors of less than $10\%$. In a different approach, we proposed a model based on high-level bit stream features \cite{Herglotz15c} such as bit rate, resolution, and number of frames. Another work showed that for software solutions, the decoding time is highly correlated to the energy \cite{Herglotz15a}. 
The more sophisticated model that we extend in this paper is presented in \cite{Herglotz13} (intraframe decoding) and \cite{Herglotz14} (P- and B-frame decoding). It is based on bit stream features and aims at modeling the energy that certain parts of the decoding process require. 

\begin{figure}
\centering
\psfrag{A}[c][c]{ENC}
\psfrag{B}[c][c]{Bit stream}
\psfrag{C}[c][c]{DEC}
\psfrag{D}[c][c]{Estimation}
\psfrag{E}[c][c]{$\hat E_\mathrm{dec}$}
\psfrag{F}[r][r]{e.g. RDO}
\psfrag{G}[l][l]{e.g. Decoder }
\psfrag{L}[l][l]{Optimization}
\psfrag{H}[c][l]{Resolution, }
\psfrag{I}[c][l]{no. frames, ...}
\psfrag{J}[l][l]{Decoding }
\psfrag{K}[l][l]{time, ...}
\includegraphics[width=0.48\textwidth]{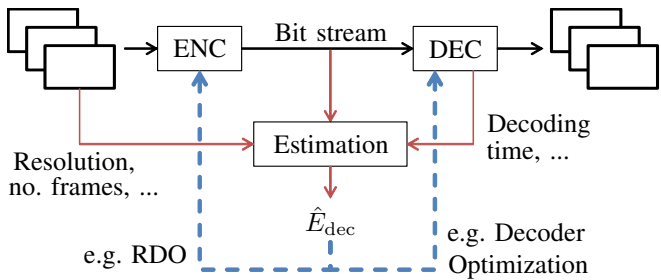}
\caption{Possible input variables (red arrows) and potential applications (blue dashed arrows) of the estimated energies. An input sequence is encoded in the encoder (ENC) and the resulting bit stream is transmitted to the decoder (DEC). The estimator can use information derived from the bit stream, from sequence specific parameters, or from decoding parameters. $\hat E_\mathrm{dec}$ is the estimated decoding energy. }
\label{fig:appl}
\end{figure}

Hence, a high amount of different models was proposed in the past to estimate decoding energy or power with the goal of reducing the power consumption on the decoder side. Typical and potential applications of these models are the following: 
\begin{itemize}
\item Real-time encoding parameter adaptation to meet decoder requirements. Due to the possibility to perform decoding energy estimations on the encoder side, the encoding parameters can be adapted to encode low energy bit streams if the end user's device reports that battery power is low. 
\item Decoding-energy optimization of video bit streams inside the Rate-Distortion-Optimization process. By incorporating the estimated decoding energy into the rate-distortion-function, energy saving bit streams could be generated targeting the decoding system of the end user. 
\item Comparison between different decoder solutions. A new algorithm can be compared to other solutions to check if it is energetically more efficient. 
\item Information about the energy consumption of different decoder modules that correspond to certain bit stream properties. This information can, e.g., be used by developers to identify the module with the highest energy consumption such that further development focuses on this module. 
\end{itemize}
Figure \ref{fig:appl} visualizes possible input variables for the energy estimator, the output energy estimation, and applications. 

In this paper, we adopt an incomplete model for HEVC-decoding and extend it such that it is able to estimate the energy consumption of any HEVC-conform bit stream. Furthermore, we compare it to several models from literature and discuss their estimation accuracies for HEVC coded sequences as well as advantages and disadvantages. Explanation on how the models can be applied will be given and an elaborate evaluation section will show the performance of the modeling approaches for different software and hardware based decoding systems. 

The rest of the paper is organized as follows: Section \ref{sec:models} introduces the energy models and describes the parameters that are used to estimate the decoding energy. A major focus is put on the proposed feature based model presented in Subsection \ref{sec:sec:feat} as descriptions in prior work \cite{Herglotz13, Herglotz14} are incomplete. Afterwards, a detailed description of the measurement setup is given in Section \ref{sec:setup}. Section \ref{sec:eval} discusses the estimation accuracy of the different models by showing their estimation performance for different decoder implementations on different hardware platforms. Finally, Section \ref{sec:concl} summarizes the conclusions of this paper.

\section{Energy Estimation Models}
\label{sec:models}
This section introduces and explains the tested models. Our goal is to show the models' energy estimation capabilities for complete decoding systems including central processing unit (CPU) power and memory access like the random access memory (RAM). Peripheral devices that are not directly related to decoding like a network connection or a display are not considered. 

The first subsection presents the proposed model based on bit stream features. In the model in Subsection \ref{sec:sec:Ren}, the energy is estimated using the number of processor events. A similar model that aims at modeling the energy consumption of the RAM is presented in Section \ref{sec:sec:ram}. We include this model in our evaluations to be able to compare the RAM-modeling capabilities of the proposed model to a dedicated reference model from literature. 

Another approach related to the properties of the decoding process is shown in Subsection \ref{sec:sec:procTime}, where the decoder processing time is used as a basis. At last, Section \ref{sec:sec:HighLevel} shows four models that are based on high-level bit stream parameters (e.g., resolution, number of frames, bit stream file size).

\subsection{Energy Model based on Bit Stream Features}
\label{sec:sec:feat}

The model presented in this subsection was first introduced for intra coding in \cite{Herglotz13} and extended to inter coded frames in \cite{Herglotz14}, where no in-loop filters were taken into consideration. In this work, we finalize this model by including the in-loop filters (the Deblocking Filter (DBF) \cite{Norkin12} and Sample Adaptive Offset (SAO) \cite{Fu12}) such that any HEVC-conform bit stream can be modeled. Specific hardware properties of the decoding system are not modeled.

The model is based on typical features that an HEVC-coded bit stream can incorporate. From an abstract point of view, these features describe which sub-processes are executed to reconstruct the sequence. Such a sub-process can, e.g., be the prediction process of one block, the residual transformation of one block, or the deblocking filtering of a block boundary. Usually, such sub-processes are implemented in high-level languages such as C++ and encapsulated in functions that are called multiple times during the decoding process. These functions show a rather constant processing flow, nevertheless their energy consumption may differ significantly upon different calls. For instance, the input data which has an impact on the number of clipping operations or the current state of the cache system can lead to a varying energy consumption. In spite of this observation, it is assumed that this variance can be neglected due to averaging when executed multiple times during the decoding process. As an advantage of this concept, concrete knowledge about the implementation of these sub-processes is not required. Hence, the first property of a bit stream feature can be described as follows: 

\begin{property}[Bit stream feature]
A bit stream feature can be associated to a sub-process during the decoding process. During the decoding of a single bit stream, these sub-processes can be executed multiple times. The sub-process consumes a specific and nearly constant processing energy $e_i$ upon each execution. 
\end{property}

In the next step, we define the set of features. A typical sub-process is, e.g., the prediction of a block of a fixed size. 
An example would be the bit stream feature ``pla($d=1$)'' (cf. Table \ref{tab:features}) which corresponds to the prediction of a block at depth $d=1$ with the planar intra-prediction mode (cf. \cite{Lainema12}). 

By such means the complete decoding process is split up into sub-processes that are associated with the bit stream features. The second important property of these features is that they can be linked to syntax elements or variables as defined in the standard \cite{ITU_HEVC}. The occurrence and value of these syntax elements and variables are used to determine how often the bit stream features occur and hence, how often the corresponding sub-processes are executed. 

\begin{property}[Bit stream feature]
By analyzing the occurrence and value of certain HEVC syntax elements or variables, it is possible to find out how often a bit stream feature occurs. The number of occurrences is denoted by the \textit{feature number} $n_i$. 
\end{property}

Hence, for a given HEVC-coded bit stream, it is counted how often a sub-process is executed. Multiplying this number with a corresponding specific energy yields the complete required decoding energy related to this sub-process. Summing up these energies for all bit stream features yields the estimated decoding energy

\begin{equation}
  \hat E_\mathrm{dec} = \sum_{\forall i} n_i \cdot e_i, 
  \label{eq:E_gen}
\end{equation}
where the index $i$ denotes the bit stream feature index, $n_i$ the feature number, %number of occurrences of the bit stream feature (feature number), 
and $e_i$ the corresponding feature-specific decoding energy. Later in Section \ref{secsec:eval} we will show how these feature-specific energies can be determined. % which is the model parameter that needs to be determined. 

The set of features is summarized in Table \ref{tab:features} in the Appendix and can be divided into five categories: 
\begin{itemize}
\item The general features (ID 1,..,3) comprise global processes (initialization of the decoding process  and the slices), 
\item Intraframe coding features (ID 4,...,10) comprise intra-prediction for different block sizes and  parsing corresponding flags, 
\item Interframe coding features (ID 11,..,24) describe inter-prediction for different block sizes, parsing, and fractional pel filtering, 
\item Residual coding features (ID 25,..,34) comprise coefficient parsing and transformation, 
\item In-loop filtering features (ID 35,..,45) comprise the in-loop filtering processes DBF and SAO.  
\end{itemize}

The feature number derivation is explained in detail in the Appendix. There, we show how feature counters implemented at positions that correspond to subclauses in the standard \cite{ITU_HEVC} can be used to derive the numbers. Using the syntax elements and variables used in the subclause, a condition can be defined that must be maintained to increment the corresponding feature number. %These subclauses and conditions are listed in detail in the Appendix. 
With the help of this list it is possible to implement the counters into any HEVC-conform decoder solution. For instance, every time a skip flag that holds the value `true' is decoded, the corresponding feature number $n_{\mathrm{skip(}d\mathrm{)}}$ at the current block's depth $d$ (which relates to the block size) is incremented.

The feature numbers can be derived by a modified decoder. A corresponding tool based on the HEVC-Test Model (HM) reference software \cite{HM-13.0} is available online \cite{denesto}. The tool incorporates counters that increment the feature numbers every time the feature occurs. To derive the feature numbers we put special attention that no reconstruction such as motion compensation or fractional pel interpolation has to be performed. Hence, depending on the input sequence, the tool requires $20\%$ to $60\%$ less processing time than the complete decoding process. In comparison to the pure decoding process, when performing simultaneous decoding and feature number derivation, the complexity of the decoder rises by $4.5\%$ in average.

As the complete set of bit stream features is rather detailed and complex, two different models are investigated as proposed in \cite{Herglotz14}. The accurate model (feature-based accurate \textit{FA}) considers the complete set of bit stream features (in total $90$ features including depth dependencies), and the simple model (feature-based simple \textit{FS}) considers a reduced set of $27$ features including depth dependencies. Table \ref{tab:features} in the Appendix indicates which features are used in which model. 
The two models are proposed for the following reasons: 
\begin{itemize}
\item The accurate model (\textit{FA}) is supposed to be used on the encoder side: When finding the best coding mode, next to the classic rate and distortion criteria, the estimated decoding energy can serve as a third decision criterion. Hence, energy saving bit streams could be constructed. For this purpose a model using a high amount of features is better suited as more information on the energetic differences between the coding modes can be exploited. 
\item The simple model can be used for applications where good estimates shall be achieved using low effort. Hence, this model is proposed for convenience and ease of application. 
\end{itemize}
Thus, the accurate model incorporates a high amount of features, though some of them may only contribute little to estimation accuracy. Further possible features like weighted prediction or the use of temporal motion vector prediction has been tested, too, but showed a very low or even negative impact on the estimation accuracy. For the simple model, all features that showed a rather low impact were discarded or merged. Note that the feature selection process for the models was rather intuitive, further work could aim at optimizing the feature set. 

The outcome of the energy estimator can be tested in an online demonstrator \cite{denesto}. It calculates the estimated decoding energy for any HEVC-conform bit stream. For further analysis, it also shows details about the distribution of the energy among the bit stream features. We encourage visiting the website and testing the functionality for different bit streams and decoder solutions.  

\subsection{Energy Model based on Processor Events}
\label{sec:sec:Ren}
The first reference model (short \textit{PE}) is proposed by Ren et al. \cite{Ren14b}. It estimates the decoding energy using processor level information. Different processor events (PEs) are counted during the decoding process and interpreted as independent variables. Afterwards, the energy is estimated using a multivariate adaptive regression spline (MARS) \cite{Friedmann91} model by  
\begin{equation}
\hat E_\mathrm{dec} = \sum_{i=0}^M \sum_{j = 0}^{n_i}  c_j \cdot B_j \left(x_i \right). 
\end{equation}
In this model, $i$ is the PE index, $M$ the number of PEs, and $x_i$ the number of occurences of the $i$th PE. For each PE, $n_i$ slopes are assigned with an associated basis function $B_j$. Ren et al. use the basis functions as a constant or a hinge function of the form 
\begin{equation}
B_j(x_i) = \max (0, x_i-k) \quad \vee \quad B_j(x_i) = \max (0, k-x_i), 
\end{equation}
where $k$ is a constant. Hence, the MARS model constructs a piecewise linear relation between the independent variables and the target function. 

Ren et al. considered the three PEs instruction fetches, level-1 data cache misses, and hardware interrupts. In our work, we only consider the former two PEs as all peripheral hardware components are disabled in the measured devices. The construction process usually returns $13$ parameters that describe the function. 

A major drawback of this model is the fact that only pure software decoder solutions can be modeled where suitable profiling tools are available. Hence, the model could only be evaluated for decoding system (a) (see Section \ref{sec:sec:decSys}), where we used Valgrind \cite{valgrind} as a profiling tool. Furthermore, the derivation of the PE numbers is highly complex due to the instruction level analysis that needs to be performed during the decoding process of the considered bit stream.

\subsection{Energy Model for the RAM}
\label{sec:sec:ram}
The model to estimate the energy consumption of the RAM is proposed in \cite{Konstantakos08} and reads
\begin{equation}
\hat E_\mathrm{dec} = e_\mathrm{ra}\cdot n_\mathrm{ra}+e_\mathrm{wa}\cdot n_\mathrm{wa} + e_\mathrm{rf}\cdot n_\mathrm{rf} + V_\mathrm{dd}\cdot i_\mathrm{ss} \cdot T_\mathrm{cp}\cdot n_\mathrm{cycles}, 
\label{eq:ram}
\end{equation}
where the parameters $e$ and the variables $n$ represent specific energies and numbers, respectively, ra stands for read-access, wa for write-access and rf for memory refresh. The last summand corresponds to the idle energy where $V_\mathrm{dd}$ is the voltage, $i_\mathrm{ss}$  the steady-state current, $T_\mathrm{cp}$ the clock period and $n_\mathrm{cycles}$ the number of clock cycles that are considered. %Hence, the model is mainly based on cache-read misses and data writes that usually have a direct influence on the energy consumption. 

As we are solely interested in the pure decoding energy we disregard the idle and refresh energy which we assume to be stationary and reduce (\ref{eq:ram}) to 
\begin{equation}
\hat E_\mathrm{dec} = e_\mathrm{ra}\cdot n_\mathrm{ra}+e_\mathrm{wa}\cdot n_\mathrm{wa}.  \label{eq:ram_simple}
\end{equation}
We obtain the number of read- and write-accesses using Valgrind \cite{valgrind}, where the number of RAM read-accesses $n_\mathrm{ra}$ (instruction and data reads) corresponds to the last-level cache misses. For the write-accesses we take the complete number of data writes as an approximation. Due to the write-back cache policy implemented in the tested system, data that are available in the cache is not always written to the RAM. A tool returning the exact number of writes was unfortunately not available for our work, hence the model only returns rough estimates for the RAM energy consumption. 
 
Note that this model does not aim at estimating decoding energies but at modeling general RAM behavior. It is included in this publication to have a reference for memory modeling. As a similar analysis as for model PE has to be performed, the derivation of the number of read- and write-accesses is highly complex and therefore only evaluated for system (a). In the following, this model is referred to by \textit{M}.

\subsection{Energy Model based on Processing Time}
\label{sec:sec:procTime}
The simplest model to estimate the energy consumption is presented in \cite{Herglotz15a}. In this paper it was shown that the processing time of the decoder is approximately linear to the decoding energy such that the energy can be estimated by 
\begin{equation}
 \hat E_\mathrm{dec} = E_\mathrm{0} + P_\mathrm{mean} \cdot t_\mathrm{dec}, 
\end{equation}
where $t_\mathrm{dec}$ is the sequence dependent processing time as, e.g., returned by the linux \texttt{time}-function or the C++ \texttt{clock()}-function. The parameter $P_\mathrm{mean}$ can be interpreted as the mean processing power and $E_\mathrm{0}$ as a constant offset. For software that is not real-time constrained this approach can return highly accurate estimates. As a drawback, estimates can only be obtained when the decoding process is executed once on the target device because the processing time needs to be measured. Furthermore, real-time constrained decoders (such as typical hardware decoders) cannot be modeled using this approach as their processing time is equal to the playback time. In the following, this model is referred to by  \textit{T}.

\subsection{Energy Models based on High-Level Parameters}
\label{sec:sec:HighLevel}
Further research has been conducted on the estimation using high-level, bit stream specific variables. To this end, Li et al. \cite{Li12} estimated the decoding power of a real time decoder using the bit stream properties frame size in pixels per frame $S$, frame rate $f$, and quantization $q$. We adapt this approach to estimate the decoding energy by 
\begin{equation}
\hat E = P_\mathrm{max} \cdot \left( \frac{S}{S\mathrm{max}}\right)^{c_S} \left( \frac{f}{f\mathrm{max}}\right)^{c_f} \left( \frac{q}{q_\mathrm{min}}\right) ^{c_q} \cdot t_\mathrm{dec}, 
\label{eq:Li}
\end{equation}
where we multiply the estimated power with the decoding time $t_\mathrm{dec}$. $S_\mathrm{max}$, $f_\mathrm{max}$, and $q_\mathrm{min}$ are the maximum and minimum values for the frame size, the frame rate, and the quantization, respectively. The exponents $c_s$, $c_f$, and $c_q$ and the maximum power $P_\mathrm{max}$ are the system specific parameters. Apparently, this model (high-level with timing \textit{H1$_T$}) is similar to the processing time based model from Section \ref{sec:sec:procTime} but incorporates more variables. Note that we are targeting general decoding systems that are not necessarily real-time constrained. Hence, the variables frame size and rate, which have a very high influence on the processing power in real-time systems, can show a much lower impact in our configuration. 

In later work, Raoufi et al. \cite{Raoufi13} identified bitrate $b$ and intra refresh rate $\alpha$ (fraction of intra coded frames of all coded frames) as suitable variables to determine the processing power. To estimate the processing energy, we again multiply with the processing time and obtain
\begin{equation}
\hat E = \left(c_1\cdot \alpha \cdot b + c_2 \cdot \alpha + c_3 \cdot b + c_4 \right) \cdot t_\mathrm{dec}, 
\label{eq:Raoufi}
\end{equation}
where $c_1 ...c_2$ are the system specific parameters. This is the second high-level feature based model using the processing time \textit{(H2$_T$)}. 

In a further step, we tried to modify the model such that it better fits the properties of an energy and not a power estimator. Therefore, we exchange the bitrate $b$ with a differently normalized variable, the mean bits per pixel $b_\mathrm{pixel}$ and multiply the result with the complete number of pixels (the product of the number of frames $N$ with the frame size $S$) and obtain
\begin{equation}
\hat E = \left(c_1\cdot \alpha \cdot b_\mathrm{pixel} + c_2 \cdot \alpha + c_3 \cdot b_\mathrm{pixel} + c_4 \right) \cdot N \cdot S. 
\label{eq:Raoufimod}
\end{equation}
As an advantage, we do not need to measure the decoding time $t_\mathrm{dec}$ on the decoding system to be able to estimate the energy. This model is referred to by \textit{H2}. 

As a last high-level model \textit{(H1)} we tested the estimation accuracy of the model presented in \cite{Herglotz15c}. It is built upon the variables bits per pixel $b_\mathrm{pixel}$, number of frames $N$ and frame size $S$ and reads 
\begin{equation}
\hat E = C + S \cdot N \cdot \left( \alpha + \beta\cdot b_\mathrm{pixel} ^\gamma\right), 
\label{eq:eusipco}
\end{equation}
where the constants $C$, $\alpha$, $\beta$, and $\gamma$ are the system specific parameters.

All the models presented in this Section are summarized in Table \ref{tab:models} including their main properties. 
\begin{table}[t]
\caption{Decoding energy models and their main properties. The parameters (par.) will be fitted to the decoding systems and the variables (var.) are fixed for a given bit stream. The last column shows if the decoding process of the bit stream has to be executed on the target device to be able to obtain energy estimates. }
\label{tab:models}
\vspace{-.4cm}
\begin{center}
\footnotesize
{\begin{tabular}{l|l|r|r|l}
\hline
Model & based on & \# par. & \# var. & Ex. required \\
\hline
(\textit{FA}) & bit stream features (accurate) & $90$ & $90$ & no \\
(\textit{FS}) & bit stream features (simple) & $27$ & $27$ &no \\
(\textit{PE}) & processor events & $\sim 13$ & $2$ &yes \\
(\textit{M}) & memory accesses &  $2$ & $2$ &yes \\
(\textit{T}) & processing time & $2$ & $1$ & yes\\
(\textit{H1$_T$}) & high-level features, time & $7$ & $4$ & yes\\
(\textit{H2$_T$}) & high-level features, time & $4$ & $3$ & yes\\
(\textit{H2}) & high-level features &$4$ & $4$ & no\\
(\textit{H3}) & high-level features & $4$ & $3$ & no\\
 \hline
\end{tabular}}
\end{center}
\end{table}

\section{Energy Measurements}
\label{sec:setup}
In order to obtain the energy required for decoding complete sequences, a dedicated test setup was constructed, capable of measuring the energy consumption of the different video decoding solutions. These measurements are used to train and validate the investigated energy models.

\begin{figure}
\centering
\psfrag{U}[c][c]{$V_0$}
\psfrag{V}[c][c]{V}
\psfrag{D}[c][c]{DEC}
\psfrag{A}[c][c]{A}
\psfrag{P}[l][c]{Power meter}
\includegraphics[width=0.4\textwidth]{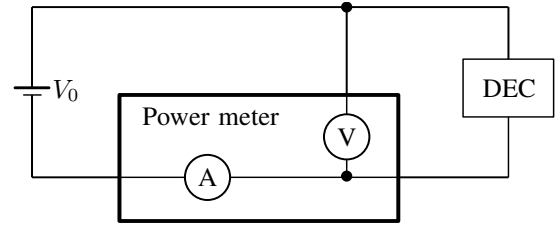}
\caption{Measurement setup with voltage source $V_\mathrm{0}$, decoding device (DEC), and power meter including an ampere meter (A) and a voltmeter (V). }
\label{fig:setup}
\end{figure}

The test setup used to measure the energy consumption of a decoding device is shown in Figure \ref{fig:setup}. 
It consists of a voltage source, a high precision power meter (ZES Zimmer's LMG95), and the decoding system (DEC). 

To obtain the power consumption of a device we measure voltage across and current through the main supply jack such that the measured energy represents the combined energy consumption of all modules (CPU, RAM, periphery, ...) on the tested board. Furthermore, two measurements were conducted to determine the energy the CPU and the RAM consume. Therefore, we identified the corresponding supply pins on the schematics of the tested board. The voltage was measured across the input and the output of the CPU and RAM power supply, respectively. The current was determined indirectly by dividing the voltage drop across the induction at the input of the power supply by the internal resistance of the induction. 

\begin{figure}
\centering
\psfrag{000}[c][c]{$0$}
\psfrag{001}[c][c]{$10$}
\psfrag{002}[c][c]{$20$}
\psfrag{003}[c][c]{$30$}
\psfrag{004}[c][c]{$40$}
\psfrag{005}[c][c]{$2.6$}
\psfrag{006}[c][c]{$2.8$}
\psfrag{007}[c][c]{$3.0$}
\psfrag{008}[c][c]{$3.2$}
\psfrag{010}[l][c]{$P_\mathrm{dec}$}
\psfrag{009}[l][c]{$P_\mathrm{idle}$}
\psfrag{011}[b][t]{Power [W]}
\psfrag{012}[t][b]{Time [s]}
\includegraphics[width=0.5\textwidth]{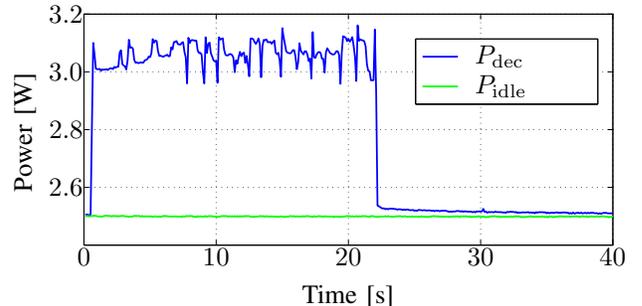}
\vspace{-1cm}
\caption{Power consumption of decoding system (a) (cf. Section \ref{sec:sec:decSys}) during the decoding process (blue) and the idle state (light green). The decoding starts at about $0.5$s and ends at $22$s. }
\label{fig:power}
\end{figure}

Figure \ref{fig:power} shows an example of the power a complete decoding system consumes (measured through the main supply jack including the main energy consumers CPU and RAM). The green curve depicts the power the system consumes in idle mode when no user process is running. It is not perfectly constant due to, e.g., memory refreshes and background processes. The blue curve shows the power consumption when a decoding process is started at about $0.5$s. Here we can see that the power rises to a higher level until the decoder finishes at about $22$s. 

If we now calculate the time-integral over the curves, we obtain the complete energy the decoding system consumes in the observed time interval. As we are only interested in the pure decoding energy $E_\mathrm{dec}$, we calculate the difference between both energies as 
\begin{equation}
	E_\mathrm{dec} = \int_{t=0}^T P_\mathrm{dec}(t)dt - \int_{t=0}^T P_\mathrm{idle}(t)dt, 
	\label{eq:Edec}
\end{equation}
where the time interval $T$ must be equal for both integrals, $P_\mathrm{dec}(t)$ is the power when the decoding system performs the decoding, and $P_\mathrm{idle}(t)$ is the power in idle mode. Visually, the resulting energy $E_\mathrm{dec}$ corresponds to the area between both curves.

To measure the energy consumption of an Intel CPU (see Section \ref{sec:sec:decSys} below) we used the Running Average Power Limit (RAPL) system \cite{David10} which has been recently proven to return very accurate approximations of the true processing energy \cite{Hackenberg15}. The energy values can be read from registers that are continuously updated. We obtain the energy of the complete decoding process by subtracting the energy value immediately before from the value immediately after the decoding process. In addition, we subtract the idle energy that we obtain by multiplying the mean processing time of the decoder with the mean idle power that we obtained in a further measurement. In our evaluation, we will use three energy values: Package which includes the energy consumption of the complete chip, Core which corresponds to the CPU cores, and Uncore which corresponds to the energy consumption of the RAM.

As the measured energies for both measurement setups are subject to different noises resulting from temperature variations, kernel processes, hardware interrupts, power meter accuracy, and other factors, we found that it is not sufficient to perform merely a single measurement for each bit stream. Hence, we perform a measurement series and incorporate a statistical test to ensure validity.

To this end, we perform a confidence interval test as proposed in \cite{Bendat71}. Consider a test sequence consisting of $m$ measurements, where we assume the samples to be normally distributed: The mean of these samples is $\bar x$, and $\sigma$ is the measurement's standard deviation. %Furthermore, we define a probability $\alpha$ that determines the width of the confidence interval. 
Given these parameters, the real mean value $\mu$ is located within probability $\alpha$ inside the interval 
\begin{equation}
	c_\mathrm{left} = \bar x - \frac{\sigma}{\sqrt{m}}\cdot t_\alpha (m-1) < \mu < 	\bar x + \frac{\sigma}{\sqrt{m}}\cdot t_\alpha (m-1) = c_\mathrm{right}, 
\label{eq:confInt}
\end{equation}
where $t_\alpha (m-1)$ denotes the critical t-value for $m$ samples.

To make sure that the true mean value is not too far from the measured mean value, we check whether the calculated confidence interval is smaller than a constant fraction of the current mean value 
\begin{equation}
	\Delta c = c_\mathrm{right}-c_\mathrm{left} = 2\cdot \frac{\sigma}{\sqrt{m}}\cdot t_\alpha (m-1)  < \beta \cdot \bar x. 
\label{eq:chosenInt}
\end{equation} 
If the condition is not satisfied, we repeat the measurement until it is met. As parameters we chose $\alpha = 99\%$ and $\beta = 0.02$. 
Doing so ensures that with a probability of at least $99\%$, the true decoding energy is not lower than $0.99$ times and not higher than $1.01$ times the measured mean energy $E_\mathrm{dec}$. Note that for the measurement on the Intel CPU the measured values are subject to much higher variation than in the other measurement setup because no background processes were switched off. Hence, for this decoding system, we drop outlying values from the calculation of the mean energy.

\section{Evaluation}
\label{sec:eval}

In this section we verify the models for several different decoding systems (Subsection \ref{sec:sec:decSys}). The data set, i.e. the bit stream set that was evaluated, is given in Section \ref{secsec:evalVids}. Section \ref{secsec:eval} presents our evaluation methodology and Section \ref{sec:anal} analyzes the estimation accuracies for all tested decoding systems and all presented estimation methods.

\subsection{Decoding Systems}
\label{sec:sec:decSys}

We set up six different test systems to validate the estimation accuracy of the models. Since one focus is on energy estimation in embedded systems, 
we decided on the Pandaboard \cite{Panda} as a representative of a typical 32-bit ARM platform used in state-of-the-art smartphones 
and tablets. Three different software solutions were tested on this device.  
Furthermore, a PC featuring an Intel CPU was investigated aiming at the energy consumption of Desktop PCs or portable Tablet PCs. For the remaining two test systems, we selected field-programmable gate array (FPGA) platforms. On the first FPGA platform a software decoder was executed on a soft intellectual property (soft IP) core to replicate the energy consumption on embedded, non-ARM CPU architectures. The second platform featured a decoder written in a hardware description language (VHDL). The results to this test series are supposed to show if the model is also valid for common, energy efficient hardware architectures like application specific integrated circuits (ASICs), which were not available for our research.

Three software decoder solutions were executed on the Pandaboard featuring Ubuntu Server 12.04 32-bit as operating system. The board represents a typical,
 embedded, multimedia platform, equipped with an OMAP4430 SoC (System-On-Chip) \cite{OMAP4430} with a 45nm feature size. The board incorporates two CPU cores on the SoC, each running at 1 GHz and a variety of additional peripheral units (wireless LAN, LCD, Camera, Universal Serial Bus (USB), HD Multimedia Interface (HDMI), 
Ethernet, etc.) that are commonly available on portable devices. The on-chip RAM is a 1GB, low power double data rate synchronous DRAM (LPDDR2). %, this board is a highly suitable candidate for our research. 
For our tests, we restricted the UNIX runlevel to 1 in order to keep the energy consumption noise of user-level processes and 
peripheral units to a minimum. The main properties of the software implementations are given in the following list: 
\begin{description}[labelsep=5pt]
\item[\normalfont{(a)}] \textbf{HM-13.0 software on the Pandaboard}: The HM software is the reference implementation of the HEVC coder. %For our measurements, we used version 13.0 \cite{HM-13.0}. 
We compiled the software on the Pandaboard using gcc for ARM with o3-optimization.
\item[\normalfont{(a$_{CPU}$)}] \textbf{CPU usage of the HM software: } The measurement of the CPU aims at the same process as in system (a). In contrast, here we only measured the energy consumption of the CPU as explained in Section \ref{sec:setup} to check if the energy consumption of this item can be modeled separately. 
\item[\normalfont{(a$_{RAM}$)}] \textbf{RAM usage of the HM software: } The same properties as for system (a$_{CPU}$), but this time the measurements were performed for the RAM. 
\item[\normalfont{(b)}] \textbf{Libde265 on Pandaboard}: This is an open-source implementation of the HEVC coder designed for real-world applications \cite{libde}. We tested version v0.7 that was compiled using gcc. 
\item[\normalfont{(c)}] \textbf{FFmpeg single core on Pandaboard}: For our third software implementation, we tested the HEVC decoder provided in the FFmpeg multimedia framework \cite{FFmpeg}, version 2.8. An interesting feature of this framework is its ability to perform multi core decoding. Hence, we split up our evaluation into single and dual core processing. For this system, the decoder is executed on a single core. 
\item[\normalfont{(c$_{dual}$)}] \textbf{FFmpeg dual core on Pandaboard}: Here we used the same configuration as with the preceding system, except that we allowed dual core decoding. Note that no voltage or frequency scaling, which helps in optimizing the energy consumption, has been tested. 
\end{description}
The Intel-based decoding system has the following properties: 
\begin{description}
\item[\normalfont{(d)}] \textbf{HM-16.6 software on Intel i7}: For this test we used a standard desktop PC featuring an Intel i7-4790 Quadcore (3.6GHz with 16GB RAM). To obtain measurements from a realistic environment, we did not switch off any background processes but only disconnected the network. The operating system was CentOS 6.X. %The measured energies are obtained using RAPL as explained above. 
Similar to system (a), we give results for the complete chipset, the CPU (d$_{CPU}$), and the RAM energy consumption (d$_{RAM}$). 
\end{description}
The main properties of the two FPGA-based systems are as follows: 
\begin{description}
\item[\normalfont{(e)}] \textbf{HM-16.5 on SPARC Leon 3 Processor}: 
This decoding system features a scalable processor architecture (SPARC) Soft IP core that is a ready-to-use hardware architecture synthesized from  VHDL. It can be modified (e.g. changes in cache size, enabling/disabling FPU (floating point unit) support, etc.) and customized at will. Afterwards, the architecture is synthesized as a custom integrated circuit (IC) or on a reconfigurable hardware like an FPGA.

The Leon3 soft IP core is a 32-bit-RISC (reduced instruction set computer) processor created by Aeroflex Gaisler \cite{gaisler2011grlib}. It is fully compliant with the SPARC-V8 instruction set. The processor was implemented onto an Altera Cyclone IV E FPGA which was mounted on a Terasic DE2-115 board at a clock frequency of $50$ MHz. For synthesis and implementation, Quartus~13 was used. All unnecessary components (e.g. FPU, memory management unit (MMU)), except for an instruction level cache, were removed. % to achieve a more precise energy estimation. 
We used the Bare-C Cross-Compiler System for Leon3 to compile the decoding software.

To remove all uncertainty regarding power estimation, no operating
system ran on this processor. Therefore, no system for submitting jobs (e.g. a
shell) is provided. For controlling purposes, we used direct hardware debugging via serial port (UART) and the debugging software GRMON2 \cite{grmon2}.
Due to file in and out (I/O) operations, the HM code cannot run directly on the Leon3; therefore, all calls were modified to use string buffers directly
compiled in the executable.

\item[\normalfont{(f)}] \textbf{Hardware implementation on Altera Stratix V FPGA}: In contrast to the aforementioned platforms, which are based on CPU cores, a completely hardwired HEVC decoder implementation was used to validate the energy modeling concept on hardware based decoders.
This decoder is based on a pure VHDL implementation, which reflects a pipeline architecture consisting of three stages: entropy decoding, motion compensation, and deblocking/SAO. The stages are coupled via coding tree unit (CTU) double buffers, guaranteeing simultaneous processing at all stages. For memory access to the frame buffer and the reference frame memory, a 32-bit wide DDR3 memory interface is used, which is not supported by a cache, as is typical for other implementations and can be found in literature.
The decoder is compatible with HM-15.0 and fully real-time capable for HEVC Main Profile @ Level 4.1. The decoder runs at a clock frequency of approx. $150$ MHz and the bit stream interface is implemented on the basis of a simple hardware ethernet stack, which only supports User Datagram Protocol (UDP) packets using a simple streaming server on a standard PC as the sender. The output of the decoder features an HDMI interface for the connection to standard monitors. Further descriptions of the decoder can be found in \cite{Engelhardt14}. 
\end{description}

\subsection{Evaluation Sequences}
\label{secsec:evalVids}
For the evaluation of our model, we encoded 10 different sequences from the HEVC test set (class A to class F) with three different quantization parameters (QPs 10, 32, and 45) and four configurations (intra, lowdelay, lowdelay\_P, and randomaccess). In this standard data set ($960$ bit streams), the low number of QPs was chosen to avoid an excessive duration of measurements. Therefore, system (a) was more thoroughly tested using an extended data set ($2880$ bit streams) with a higher number of QPs ranging from 5 to 50 in steps of 5. For all sets, one to eight frames were coded for each sequence. The sequences and their main properties are listed in Table \ref{tab:eval_vids}.

\begin{table}[t]
\renewcommand{\arraystretch}{1.3}
\caption{Properties of standard evaluation data set. The sequences were taken from the HEVC test set and encoded with HM-13.0 using the configurations intra, lowdelay\_P, lowdelay, and randomaccess. QP was set to 10, 32, and 45.  }
\label{tab:eval_vids}
\vspace{-.4cm}
\begin{center}
\begin{tabular}{l|c|c}
\hline
Name & Class & Resolution  \\
\hline
PeopleOnStreet & A & $2560\times1600$  \\
Traffic & A & $2560\times1600$ \\
Kimono & B & $1920\times1080$ \\
RaceHorses & C & $832\times480$ \\
BasketballPass & D & $416\times240$ \\
BlowingBubbles & D & $416\times240$ \\
BQSquare & D & $416\times240$ \\
RaceHorses & D & $416\times240$ \\
vidyo3 & E & $1280\times720$ \\
SlideEditing & F & $1280\times720$ \\
 \hline
\end{tabular}
\end{center}
%\vspace{-.5cm}
\end{table}

For the measurement of the Leon3 on the Altera Board (e), only sequences taken from classes B, C, D, and F were evaluated, due to the relatively small size of the RAM ($613$ bit streams). Furthermore, a different set of test sequences was evaluated for the hardware implementation (f) as it features some architectural constraints ($412$ bit streams). One of them is the fact that it can only decode and display sequences with HD-resolution ($1920\times1080$ pixels). Besides, a QP lower than 20 and, depending on the content, QPs up to 35 could not be tested as the resulting bit rate was too high for the implemented entropy decoding engine. The list of sequences is given in Table \ref{tab:eval_vids_HHI_HEVC}. 
\begin{table}[t]
\renewcommand{\arraystretch}{1.3}
\caption{Properties of evaluation data set for the hardware implementation (system (f)). The sequences were taken from the HEVC test set and were encoded with HM-13.0. QP was set in the range from $25$ to $50$. }
\label{tab:eval_vids_HHI_HEVC}
\vspace{-.4cm}
\begin{center}
\footnotesize{\begin{tabular}{l|c|c}
\hline
Name & Class & Resolution  \\
\hline
BasketballDrive & B & $1920\times1080$ \\
BQTerrace & B & $1920\times1080$  \\
Cactus & B & $1920\times1080$ \\
Kimono & B & $1920\times1080$  \\
ParkScene & B & $1920\times1080$  \\
 \hline
\end{tabular}}
\end{center}
\vspace{-.5cm}
\end{table}

\subsection{Evaluation Methodology}
\label{secsec:eval}

As a quality metric to rate the estimation performance of the models we choose the relative estimation error as 
\begin{equation}
\varepsilon =  \frac{\hat E_\mathrm{dec}-E_\mathrm{dec}}{E_\mathrm{dec}}, 
\label{eq:eps}
\end{equation}
where $\hat E_\mathrm{dec}$ is the estimated and $E_\mathrm{dec}$ the measured (true) decoding energy. This approach returns more meaningful results than using the absolute error as we aim at estimating the decoding energy accurately independent from the sequence resolution and length. 
Then, we average the estimation error for a given decoding system and a given model over the measured bit stream set by 
\begin{equation}
\overline{\varepsilon} = \frac{1}{M} \sum_{m=1}^M  \left| \varepsilon_m\right| , 
\label{eq:eps_mean}
\end{equation}
where $m$ is the bit stream index and $M$ the total number of evaluated streams in the corresponding data set. By averaging over the absolute value of the relative error we obtain the mean value of the deviations. 

\begin{figure}[t]
\centering
\psfrag{A}[c][c]{Coded Evaluation Sequences (Table \ref{tab:eval_vids})}
\psfrag{B}[c][c]{Bit Stream Analysis}
\psfrag{C}[c][c]{Measurement}
\psfrag{D}[c][c]{Bit stream specific variables}
\psfrag{E}[c][c]{Decoding energies $E_\mathrm{dec}$}
\psfrag{F}[c][c]{Training}
\psfrag{G}[c][c]{Validation}
\psfrag{H}[c][c]{Model}
\psfrag{I}[c][c]{parameters}
\psfrag{J}[l][l]{$10$-fold cross-validation}
\psfrag{K}[c][c]{Estimation error $\overline{\varepsilon}$}
\includegraphics[width=0.47\textwidth]{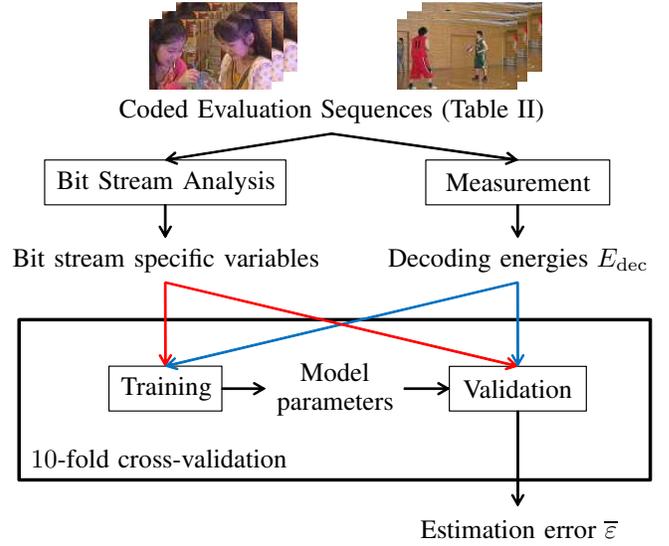}
\caption{Evaluation flow. The coded evaluation sequences (cp. Section \ref{secsec:evalVids}) are analyzed for bit stream specific variables like feature numbers, high-level features or decoding times as introduced in Section \ref{sec:models}. Furthermore, the decoding energy of these sequences is measured for all decoding systems (cf. Section \ref{sec:sec:decSys}). Afterwards, for each model and each decoding system we feed the bit stream specific variables and the decoding energies into a 10-fold cross-validation loop to train the model parameters and validate the estimation accuracy. As an output, we obtain the mean absolute estimation errors $\overline{\varepsilon}$ (cf. (\ref{eq:eps_mean})) for the complete data set.   }
\label{fig:eval_flow}
\end{figure}

All the steps that are performed to obtain these errors are summarized in Figure \ref{fig:eval_flow}. In the first step, to determine the model parameter values (i.e. the specific energies), for each decoding system and each model we perform a least-squares fit using a trust-region-reflective algorithm as presented in \cite{Coleman96}. As an input we use the measured energies for a subset of the sequences (the training set) and their corresponding variables, i.e. the sequence dependent feature numbers. As an output we obtain the least-squares optimal parameters (i.e. the optimal specific energies) for the input training set, where we train the parameters such that the mean relative error as shown in (\ref{eq:eps_mean}) is minimized. These model parameters are then used to validate the accuracy of the model on the remaining validation sequences. %The processing flow of this training is depicted in the black box in Figure \ref{fig:eval_flow}. 

The training and validation data set is determined using a $10$-fold cross-validation as proposed in \cite{Zaki14}. In this technique, we randomly divide the complete set of measured energies into $10$ approximately equally sized subsets. Then, for each subset, we use the other $9$ subsets to train the models as explained above. The trained parameters are then used to validate the remaining subset by calculating the relative estimation errors (cf. (\ref{eq:eps})) for all sequences.

\begin{table*}[ht]
\renewcommand{\arraystretch}{1.3}
\caption{Mean estimation errors for decoding system (a) and all tested models. The mean relative errors $\overline\varepsilon$ are given in percentage. The right column gives the average row-wise estimation errors and the bottom row the average column-wise estimation errors.  }
\label{tab:errors_standard}
\vspace{-.4cm}
\begin{center}
\footnotesize{\begin{tabular}{l||c|c|c|c|c|c|c|c|c||c}
\hline
 &\textit{FA} & \textit{FS} & \textit{PE} & \textit{M} & \textit{T} & \textit{H1$_T$} & \textit{H2$_T$} & \textit{H2} & \textit{H3}  & $\varnothing$ \\
 \hline
(a) & $6.21\%$  & $6.27\%$ & $8.58\%$ & $8.86\%$ & $6.04\%$ & $5.96\%$ & $6.13\%$ & $20.30\%$ & $14.74\%$ & $9.23\%$\\
%(a) & $\%$  & $\%$ & $\%$  & $\%$ & $\%$ & $\%$ & $\%$ & $\%$ \\
(a$_{CPU}$)  & $6.41\%$  & $6.77\%$ & $9.30\%$  & $10.82\%$ & $7.30\%$  & $6.93\%$ & $7.90\%$ & $21.76\%$ & $15.72\%$ & $10.32\%$\\
(a$_{RAM}$)  & $16.20\%$  & $15.80\%$ & $17.55\%$  & $18.54\%$ & $37.82\%$ & $31.38\%$ & $20.04\%$ & $23.57\%$ & $24.19\%$ & $22.79\%$\\
\hline
$\varnothing$  & $9.61\%$ & $9.61\%$ & $11.81\%$ & $12.74\%$ & $17.05\%$ & $14.76\%$ & $11.36\%$ & $21.88\%$ &  $18.22\%$\\
 \hline
\end{tabular}}
\end{center}
\end{table*}

During our investigations we observed that the estimation accuracy of the models highly depends on the used video sequences. Particularly, we noticed that high-resolution sequences with a large number of frames are estimated much more accurately than short sequences with a low resolution. 
Hence, to make sure that comparisons performed in this section are fair, we only compare estimation errors for the exact same set of measured video data. We perform the analysis on frame level as all presented models perform sufficiently well for large sequences. Frame-level values are obtained by subtracting the measured energy of a sequence containing $n-1$ frames from the energy of the corresponding sequence containing $n$ frames. Note that as a consequence, reported estimation errors from literature may differ significantly from those presented in this publication.

As shown in Section \ref{secsec:evalVids}, we used different data sets for different decoding solutions. Hence, it is not possible to directly compare the estimation error for different decoding systems. As a counter measure we evaluate the decoding process of each data set on system (a) such that the relative estimation performance can be compared to this system.

\subsection{Discussion}
\label{sec:anal}

Due to the high amount of data we split the discussion of the results into three parts. First, we compare the estimation errors of the different models based on decoding system (a) to point out their strengths and weaknesses. Second, we talk about the different decoding systems and how accurately they can be modeled by the given approaches. Third, further observations are summarized.

\subsubsection{Model Performance}

The estimation errors of all models for system (a) are summarized in Table \ref{tab:errors_standard}. 
Looking at the average estimation errors of models \textit{FA} and \textit{FS} we can see that in general, they perform best (average errors smaller than $10\%$), which can be explained by the high amount of parameters. As both models show a similar estimation error for all decoding systems and model \textit{FS} requires fewer modeling parameters, we will only evaluate model \textit{FS} in the following. 

Generally, all the models using the processing time (\textit{T, H1$_T$, } and \textit{H2$_T$}) return similar results for all decoding systems. Sound estimations (around $6\%$ which is close to the errors of models \textit{FA} and \textit{FS}) are especially obtained for decoding device (a). In contrast, modeling the RAM of system (a) only works well with model \textit{H2$_T$} ($20.04\%$), which can be explained by the fact that this model takes into account the fraction of intra coded frames which has a direct influence on the memory usage (inter frames require more memory access to reference frames). For the RAM, the other decoding time based models fail which can be explained by, e.g., the influence of the RAM's unknown initial state. Hence, in the following, we will only evaluate model \textit{H$2_T$} as in general, it returns the best results. 

Model \textit{PE} returns higher estimation errors than the decoding time based models ($8\%$ in comparison to $6\%$ for system (a)). As it additionally requires a very complex profiling on the decoding system, the decoding time based approaches seem to be more convenient in practical use. 
The values for model \textit{M} indicate that even a dedicated memory model does not estimate the memory energy more accurately ($18.54\%$) than the proposed models \textit{FA} and \textit{FS} (around $16\%$) which can be explained by the simplicity of the model. Furthermore, the memory variables (reads and writes) seem to have a very high correlation with the complete energy as estimation errors for the systems (a) and (a$_{CPU}$) are below $11\%$. 

Considering the high-level models \textit{H2} and \textit{H3}, the estimation errors are significantly higher than the other models ($15\%-20\%$ in comparison to less than $9\%$ for system (a)). Furthermore, model \textit{H3} returns better results than model \textit{H2} (around $5\%$ lower estimation error).

\subsubsection{Decoding Systems}
Table \ref{tab:allSystems} gives the mean estimation errors for decoding systems (b) to (d). 
\begin{table}[t]
\renewcommand{\arraystretch}{1.3}
\caption{Mean estimation errors for decoding systems (b) to (d) and a representative subset of the tested models. Note that models \textit{PE} and \textit{M} could not be evaluated for these systems. }
\label{tab:allSystems}
\vspace{-.4cm}
\begin{center}
\footnotesize{\begin{tabular}{l||c|c|c|c}
\hline
& \textit{FS} & \textit{H2$_T$} & \textit{H2} & \textit{H3} \\% & $\varnothing$\\
 \hline
(b)  & $4.49\%$ & $9.30\%$ & $29.06\%$ & $22.46\%$ \\%& $12.18\%$\\
(c)  & $6.77\%$ & $14.79\%$ & $20.93\%$ & $16.95\%$ \\%& $14.45\%$\\
(c$_{dual}$) & $7.48\%$ & $24.67\%$ & $21.03\%$ & $16.87\%$ \\%& $18.38\%$\\
(d) & $12.83\%$ & $45.09\%$ & $24.69\%$ & $19.74\%$\\% & $29.57\%$ \\
(d$_{CPU}$) & $9.00\%$ & $35.86\%$ & $22.89\%$ & $16.20\%$ \\%& $23.60\%$ \\
(d$_{RAM}$) & $50.54\%$ & $102.29\%$ & $99.25\%$ & $62.65\%$ \\%& $82.67\%$ \\
%\hline
%$\varnothing$ &$13.33\%$ & $33.61\%$ & $31.50\%$ & $23.28\%$ & \\
 \hline
\end{tabular}}
\end{center}
\end{table}
First of all, considering the Pandaboard decoders (b) and (c), we can see that their estimation errors are comparable to the values of (a). In contrast, the time-based model \textit{H2$_T$} fails for system (c) (value greater than $14\%$). Furthermore, we can see that the high-level models \textit{H$_2$} and \textit{H$_3$} are more accurate in modeling the Intel system than the decoding time based model \textit{H2$_T$} (around $20\%$ in comparison to more than $30\%$), which indicates that processing time is an unsuitable predictor for complex instruction set (CISC) architectures. 

Considering the energy consumption for CPU and RAM of systems (a) and (d) as shown in Tables \ref{tab:errors_standard} and \ref{tab:allSystems}, we can see that the CPU estimates are sound. In contrast, the estimation of the RAM is weak as its error is higher than $10\%$ and $50\%$, respectively. Apparently, the influence on the combined error of (a) and (d) is rather low which can be explained by the RAM occupation which is less than $10\%$ of the complete energy consumption for both systems, as indicated by our measurements. Please note that as stated in Section \ref{sec:setup}, we only consider the additional processing energy, the RAM occupation of the complete energy is much higher when the idle energies of the RAM and the CPU are included.

The estimation errors for the FPGA-based systems (e) and (f) are given in Tables \ref{tab:errors_sparc} and \ref{tab:errors_HHI}, respectively. 
\begin{table}[t]
\renewcommand{\arraystretch}{1.3}
\caption{Mean estimation errors for decoding systems (a) and (e). The data set consists of the sequences from Table \ref{tab:eval_vids} that were small enough to be decoded on the FPGA.   }
\label{tab:errors_sparc}
\vspace{-.4cm}
\begin{center}
\footnotesize{\begin{tabular}{l||c|c|c|c}
\hline
 & \textit{FS} &  \textit{H2$_T$} & \textit{H2} & \textit{H3}\\% & $\varnothing$ \\
 \hline
(a)  & $6.69\%$ & $7.38\%$ & $45.75\%$ & $42.20\%$\\% & $17.13\%$\\
(e) & $6.07\%$ & $8.77\%$ & $59.10\%$ & $34.09\%$ \\%& $11.38\%$\\
 \hline
\end{tabular}}
\end{center}
\end{table}
\begin{table}[t]
\renewcommand{\arraystretch}{1.3}
\caption{Mean estimation errors for decoding systems (a) and (f). The video data set consists of the HD sequences from Table \ref{tab:eval_vids_HHI_HEVC}. }
\vspace{-.4cm}
\label{tab:errors_HHI}
\begin{center}
\footnotesize{\begin{tabular}{l||c|c|c}
\hline
 & \textit{FS} & \textit{H2} & \textit{H3} \\%& $\varnothing$ \\
 \hline
(a) & $6.38\%$ & $14.55\%$  & $11.00\%$\\%& $9.64\%$\\
(f) & $14.89\%$ & $18.32\%$  & $27.51\%$\\% & $19.99\%$\\
 \hline
\end{tabular}}
\end{center}
\end{table}
For system (e) we can see that decoding on the SPARC returns similar estimation errors as system (a). The high-level models perform worse in this case because only low-resolution sequences were tested.  Estimating the hardware decoder (f) returns rather high estimation errors (more than $14\%$). This observation has two reasons: First, the fraction of the RAM on the complete power consumption is higher than for the software based systems (about $30\%$ as indicated by measurements using the tool Power Monitor \cite{StratixVManual}). Second, the structure of such an implementation is completely different from a software decoder %. E.g., software based systems execute certain functions only when needed, hardware based systems often do not implement a corresponding branch such that these functions may be executed independent from the content of the sequence. Still we can see that the proposed model achieves the lowest estimation errors, but a 
such that a different model that respects the different characteristics may be more suitable. Unexpectedly, the high-level model \textit{H$_2$} performs relatively well ($18\%$ estimation error) which indicates that that the fraction of intra coded frames is an important factor in the energy consumption of this hardware implementation. 

\subsubsection{Further Observations}
To show that the tested models are independent from the QP, we tested system (a) for a higher number of different QPs (QPs $5$ to $50$ in steps of $5$). The resulting estimation errors turned out to be very close to the values given in Table \ref{tab:errors_standard} such that our evaluation method using only three QP values is confirmed. 

Next, we would like to discuss the results shown in Table \ref{tab:errors_longSeq}. 
\begin{table}[t]
\renewcommand{\arraystretch}{1.3}
\caption{Mean estimation errors for a subset of $120$ bit streams from the standard test set containing eight frames (cf. Table \ref{tab:eval_vids}). We can see that in comparison to Table \ref{tab:errors_standard}, the estimation errors decrease significantly.  }
\label{tab:errors_longSeq}
\vspace{-.4cm}
\begin{center}
\footnotesize{\begin{tabular}{l||c|c|c|c}
\hline
 & \textit{FS} &  \textit{H2$_T$} & \textit{H2} & \textit{H3}\\%  & $\varnothing$ \\
 \hline
(a)   & $1.39\%$ & $2.40\%$ & $16.65\%$ & $15.00\%$ \\%& $5.76\%$\\
(a$_{CPU}$) & $2.91\%$ & $7.34\%$ & $18.49\%$ & $14.54\%$\\% & $7.69\%$\\
(a$_{RAM}$)  & $3.47\%$ & $9.91\%$ & $14.92\%$ & $10.50\%$ \\%& $14.04\%$\\
(d) & $2.61\%$ & $13.49\%$ & $21.07\%$ & $15.19\%$ \\%& $10.53\%$ \\
(d$_{CPU}$) & $2.94\%$ & $10.31\%$ & $19.64\%$ & $16.61\%$ \\%& $10.06\%$\\
(d$_{RAM}$)  & $4.91\%$ & $33.90\%$ & $22.58\%$ & $8.96\%$ \\%& $24.30\%$\\
 \hline
\end{tabular}}
\end{center}
\end{table}
Here, we can see how the estimation errors change when using sequences containing multiple frames. First, we can observe that the estimation errors drop significantly (in average the error is halved for all tested models and decoding systems). %, also for the models that are not listed). 
This can be explained by the law of large numbers where estimation inaccuracies are averaged. Particularly, we can see that sound estimations can even be obtained for the RAM (model \textit{FS} lower than $5\%$). For the hardware implementation (system (f)) the mean error does not change significantly in comparison to the single frame approach. 

Finally, to show that our results are consistent, we would like to discuss reported values from literature. Basically, these values are close to the values given in Table \ref{tab:errors_longSeq} that correspond to multiple frame sequences (e.g., models \textit{FS, T, H$_3$}, \textit{H1$_T$}). Intuitively, this corresponds to our expectation as in related work, only sequences containing multiple frames were tested. For the feature based models, we could even achieve a lower estimation error ($3.63\%$ in \cite{Herglotz14}) which is caused by the more sophisticated training method. In addition, reported errors for model \textit{PE} correspond to our findings (more than three quarters of the tested sequences in \cite{Ren14b} showed an estimation error between $5-10\%$ where we found a mean error of $\sim 8\%$). In contrast, Raoufi et al. reported estimation errors of less than $1\%$ (models \textit{H2} and \textit{H2$_T$}), which is much better than our results. However, they used an H.264 decoder and only tested a single sequence such that a lower error can be expected.

\section{Conclusion}
\label{sec:concl}

In this paper we proposed a bit stream feature based model and revisited some models capable of estimating the required decoding energy for a given HEVC-coded bit stream. We tested the models meticulously for different software solutions, one hardware implementation, and processing devices with a large amount of test sequences to obtain meaningful estimation errors. As a result we have seen that best results can be obtained using the proposed simple feature based model using $20$ parameters and that software decoder solutions can be best modeled. Accepting estimation errors of up to $20\%$ the model can also be applied to hardware based decoders. 

In future work, we plan to validate, test, and adapt the model for specific hardware decoders like ASICs. Furthermore, we will utilize the feature based model inside the Rate-Distortion-Optimization process to encode sequences such that the decoding energy can be minimized. Another application would be to provide a real time task scheduler with the gathered information. As the energy estimation can be calculated before the particular decoding takes place, for instance in a pipelined decoder structure, sophisticated power saving strategies can be applied to software and hardware based decoder systems with the presented results. %Another interesting direction 

\appendix[The Bit Stream Feature Set for the Feature Based Model]
The bit stream features used in models \textit{FA} and \textit{FS} are listed in Table \ref{tab:features}. The table indicates how the feature numbers can be derived and to which model they belong. 

The feature label helps to elucidate the meaning of the feature. The column ``Depths'' gives information about the depths at which the feature can occur. 
The columns ``Subclause'', ``Condition for syntax element or variable'', and ``Inc'' provide information about the derivation of the feature numbers. The last two columns ``\textit{FS}'' and ``\textit{FA}'' indicate whether the feature is used for the simple or the accurate model (see Section \ref{sec:sec:feat}), respectively.

First, the column ``Depths'' in Table \ref{tab:features} specifies whether the feature numbers are counted for the individual depths of the quadtree decomposition. In this context, depth 0 stands for a $64 \times 64$ block and goes down to a block size of $4\times 4$ at depth 4.  ``all'' specifies that the feature is counted at every depth. 

The following  explanation is directly related to the Recommendation ITU-T H.265 \cite{ITU_HEVC}. 
The column labeled ``Subclause'' specifies at which point the counter of the feature number can be placed during the parsing process, where the subclause refers to the subclause in the Recommendation. Furthermore, the 5th column ``Condition for syntax element or variable'' describes the syntax element or variable that must be checked to determine whether the number shall be increased. % at a certain point. 

If there is a checkmark in column ``Inc'', the feature number can simply be incremented so long as the condition described in the 5th column is met. For instance, the feature number $n_\mathrm{coeff}$ is incremented every time the \texttt{significant\_coeff\_flag} is set during the parsing process described in Subclause 7.4.9.11 in the Recommendation \cite{ITU_HEVC}. 

The remaining feature numbers without a checkmark in column ``Inc'' are derived as follows: 
\begin{itemize}
\item $E_\mathrm{0}$: As this feature describes the initialization of the decoding process, it always occurs once for each bit stream. Hence, its number is constantly $n_{E_\mathrm{0}} = 1$. 

\item fracpelVer: This feature describes the filtering process of one fractional luma pel in vertical direction. Each time the given condition (the $\%$-sign denotes the modulo-operator) is met during the parsing of the given subclause, the number $n_\mathrm{fracpelVer(d)}$ at the current depth d has to be increased by the number of luma pels to be filtered. In terms of the variables used in the standard the equation reads 
\begin{equation}
	n_\mathrm{fracpelVer(d)} = n_\mathrm{fracpelVer(d)} + \mathtt{nPbW} \cdot \mathtt{nPbH},  
	\label{eq:fracVer}
\end{equation}
where $\mathtt{nPbW}$ describes the number of luma pels in a horizontal and $\mathtt{nPbH}$ in a vertical orientation in the current prediction unit (PU).

\item fracpelHor:  To derive this number, the same equation holds as for vertical filtering: 
\begin{equation}
	n_\mathrm{fracpelHor(d)} = n_\mathrm{fracpelHor(d)} + \mathtt{nPbW} \cdot \mathtt{nPbH},  
	\label{eq:fracHor_1}
\end{equation}
Note that any filtering performed in both vertical and horizontal directions (denoted by the condition in brackets) must be checked. In this case, further horizontal filtering operations must be considered as for the subsequent vertical filtering, horizontally filtered fractional pels outside of the current PU have to be calculated. We consider this fact by 
\begin{equation}
	n_\mathrm{fracpelHor(d)} = n_\mathrm{fracpelHor(d)} + 6 \cdot \mathtt{nPbW},  
	\label{eq:fracHor_2}
\end{equation}
as six additional rows of fractional pels are required, three above and three below the current PU. 

\item fracpelAvg: This feature number comprises all horizontal and vertical luma pel filterings: %is the sum of the numbers $n_\mathrm{fracpelVer}$ and $n_\mathrm{fracpelHor}$. 
\begin{equation}
	n_\mathrm{fracpelAvg} = \sum_{\mathrm{d}=0}^3 n_\mathrm{fracpelVer(d)} + n_\mathrm{fracpelHor(d)}.   
	\label{eq:fracAvg}
\end{equation}

\item chrHalfpel: As the chroma pels are subsampled, we must consider the case in which the motion vector points to a chroma-half pel position (which is a full pel position for luma samples). Thus, each time the condition is true we apply the following equation: 
\begin{equation}
	n_\mathrm{chrHalfpel(d)} = n_\mathrm{chrHalfpel(d)} + \frac{\mathtt{nPbW}}{2} \cdot \frac{\mathtt{nPbH}}{2}.   
	\label{eq:chrHalfpel}
\end{equation}
Evaluations showed that a further decomposition into horizontal and vertical direction as well as the calculation of outlying pels has a minor impact on the estimation accuracy so that we neglect these possibilities.

\item bi: This feature number counts the number of bipredicted $4\times 4$ blocks. If the condition is met, it is calculated as 
\begin{equation}
	n_\mathrm{bi} = n_\mathrm{bi} + \frac{\mathtt{nPbW}}{4} \cdot \frac{\mathtt{nPbH}}{4}.   
	\label{eq:bi}
\end{equation}

\item MVD (motion vector difference): If the given condition is met, this feature number is increased by 
\begin{equation}
	n_\mathrm{MVD} = n_\mathrm{MVD} + \log_2\left( \mathtt{abs\_mvd\_minus2}+2 \right). 
	\label{eq:MVD}
\end{equation}
The logarithm of the value was chosen due to the exponential properties of the Exp-Golomb-Code that was used to code the motion vector difference \cite{Sze12}. The real parsing energy is assumed to be nearly linear to this expression.\footnote{Equation (\ref{eq:MVD}) can be approximated using fixed point arithmetics. In our implementation we chose to use the position of the highest nonzero bit of $\mathtt{abs\_mvd\_minus2}+2$ to approximate the logarithm. Furthermore, if the bit next to the highest bit is also set, we add a refining value of $\log_2\left(1.5\right) = 0.585$. }

\item val: The residual coefficient value is coded in the same way as the MVD. Hence, the equation reads
\begin{equation}
	n_\mathrm{val} = n_\mathrm{val} + \log_2\left( \mathtt{coeff\_abs\_level\_remaining}+2 \right). 
	\label{eq:val}
\end{equation}

\item Tr: This feature comprises all transformation processes depending on the transform size. Hence, depth 1 relates to a $32\times 32$ transform, depth 2 to a $16\times 16$ transform, and so on. Hence, if a coded block flag (CBF) is set for one of the chroma components, the counter has to be incremented for depth $\mathrm{d}=min(\mathrm{d}+1,4)$. Thus, the smaller transform block size for chroma blocks is taken into account. 

\item SAO\_allComps: As a final feature number $n_\mathrm{SAO\_allComps}$ is incremented if all color components are filtered using SAO. Measurements showed that in this case less energy is required in contrast to the processing of single color components which is reflected in a negative value of the specific energy. 
\end{itemize}

\begin{table*}[h]
%{\small
%\hfill{}
%\renewcommand{\arraystretch}{1.3}
\caption{Bit stream features used to estimate the energy consumption of a video decoder. The first column shows the feature ID, the second column its label. If the label is followed by (d), the feature is depth dependent and defined for the depths shown in column ``Depths''. The next three columns (``Subclause'' to ``Inc'') give information about the position and the condition of the corresponding feature number counters. The checkmark in the columns \textit{FS} and \textit{FA} state if the feature is used in the simple and accurate feature based model, respectively. }
\label{tab:features}
%\begin{center}
\centering
\begin{tabular}{r|l|c|c|c|c|c|c}%|r }
\hline
ID & Feature label & Depths & Subclause & Condition for syntax element or variable & Inc & \textit{FS} & \textit{FA} \\% & $e_f$ [J] \\
\hline
1 & $E_\mathrm{0}$ & - &- & - & - & $\surd $ &$\surd $ \\%& $1.111111 $\\
2 & Islice & - & 7.4.7.1 & \texttt{slice\_type == I} & $\surd$ &$\surd $ & $\surd $ \\%& $1.111111 $\\
3 & PBslice & - & 7.4.7.1 &\texttt{slice\_type == P || slice\_type == B} & $\surd$ & $\surd $ &$\surd $  \\%& $2.111111 $\\
\hline
4 & intraCU & all & 7.4.9.5 & \texttt{pred\_mode\_flag == $1$} & $\surd$ & $\surd$ & $\surd$  \\%& $0.02$ \\
5 & pla(d)  &$1..4$ & 8.4.4.2.1 & \texttt{IntraPredModeY == $0$} & $\surd$ & - & $\surd $ \\%& $2.111111 $\\
6 & dc(d)  &$1..4$ & 8.4.4.2.1 & \texttt{IntraPredModeY == $1$} & $\surd$ & - & $\surd $  \\%& $2.111111 $\\
7 & hvd(d)  &$1..4$ & 8.4.4.2.1 & \texttt{IntraPredModeY $\in \{2,10,26,34\}$} & $\surd$ & - & $\surd $ \\%& $2.111111 $\\
8 & ang(d)  &$1..4$ & 8.4.4.2.1 & \texttt{IntraPredModeY} $\in \{3,..,33\} \backslash \{10,26\}$& $\surd$ & - & $\surd $ \\% & $2.111111 $\\
9 & all(d)  &$1..4$ & 8.4.4.2.1 & always & $\surd$ &$\surd $ & -  \\%& $2.111111 $\\
10 & noMPM & all & 7.4.9.5 & \texttt{prev\_intra\_luma\_pred\_flag == $0$} & $\surd$ & - & $\surd$ \\%& $0.02$ \\
\hline
11 & skip(d) &  $0..3$ & 7.4.9.5 & \texttt{cu\_skip\_flag == $1$} & $\surd$ & $\surd$ & $\surd$ \\%& $0.02$ \\
12 & merge(d) &  $0..3$ & 7.4.9.6 & \texttt{merge\_flag == $1$ \&\& PartMode == $0$} & $\surd$ & - & $\surd$\\% &   $0.02$ \\
13 & mergeSMP(d) &  $0..3$ & 7.4.9.6 & \texttt{merge\_flag == $1$ \&\& PartMode $\in \{1,2\}$} & $\surd$ & - & $\surd$\\% & $0.02$ \\
14 & mergeAMP(d) &  $0..2$ & 7.4.9.6 & \texttt{merge\_flag == $1$ \&\& PartMode $\in \{4,5,6,7\}$} & $\surd$ & - & $\surd$\\% &  $0.02$ \\
15 & inter(d) &  $0..3$ & 7.4.9.6 & \texttt{merge\_flag == $0$ \&\& PartMode == $0$} & $\surd$ & - & $\surd$ \\%&   $0.02$ \\
16 & interSMP(d) &  $0..3$ & 7.4.9.6 & \texttt{merge\_flag == $0$ \&\& PartMode $\in \{1,2\}$} & $\surd$ & - & $\surd$ \\%& $0.02$ \\
17 & interAMP(d) &  $0..2$ & 7.4.9.6 & \texttt{merge\_flag == $0$ \&\& PartMode $\in \{4,5,6,7\}$} & $\surd$ & - & $\surd$\\% &   $0.02$ \\
18 & interCU(d) &  $0..3$ & 7.4.9.5 & \texttt{CuPredMode != MODE\_INTRA \&\& cu\_skip\_flag == $0$} & $\surd$ & $\surd$ & -\\% & $0.02$ \\
19 & fracpelHor(d) &  $0..3$ & 8.5.3.3.3.1 & \texttt{mvLX[0] \% $4$ != $0$}  (\texttt{\&\& mvLX[1] \% $4$ != $0$}) & - & - & $\surd$ \\%& $0.02$ \\
20 & fracpelVer(d) &  $0..3$ & 8.5.3.3.3.1 & \texttt{mvLX[1] \% $4$ != $0$} & - & - & $\surd$  \\%& $0.02$ \\
21 & fracpelAvg & all & 8.5.3.3.3.1 & \texttt{mvLX[0] \% $4$ != $0$ || mvLX[1] \% $4$ != $0$} & - & $\surd$ & -\\% &   $0.02$ \\
22 & chrHalfpel(d) &  $0..3$ & 8.5.3.3.3.1 & \texttt{mvLX[0] \% $8$ == $4$; mvLX[1] \% $8$ == $4$} & - & - & $\surd$\\%& $0.02$ \\
23 & bi & all & 8.5.3.2 & \texttt{predFlagL0 == predFlagL1 == $1$} & - & $\surd$ & $\surd$ \\%& $0.02$ \\
24 & MVD & all & 7.4.9.9 & \texttt{abs\_mvd\_greater1\_flag == $1$} & - & - & $\surd$ \\%& $0.02$ \\
\hline
25 & coeff & all & 7.4.9.11 & \texttt{significant\_coeff\_flag == $1$} & $\surd$ & $\surd$ & $\surd$\\% & $0.02$ \\
26 & coeffg1 & all & 7.4.9.11 & \texttt{coeff\_abs\_level\_greater1\_flag == $1$} & $\surd$ & - & $\surd$ \\%& $0.02$ \\
27 & CSBF & all & 7.4.9.11 & \texttt{coded\_sub\_block\_flag == $1$ \&\& ( xS, yS )!=( 0, 0 )} & $\surd$ & - & $\surd$ \\%& $0.02$ \\
28 & val & all & 7.4.9.11 & \texttt{coeff\_abs\_level\_remaining}%\texttt{$\log_2\left(\mathtt{coeff\_abs\_level\_remaining}+2\right)$} 
& - & $\surd$ & $\surd$ \\%& $0.02$ \\
29 & TrIntraY(d) & $1..4$ & 7.3.8.10 & \texttt{CuPredMode == MODE\_INTRA \&\& cbf\_luma == $1$} & $\surd$ & - & $\surd$ \\%&  $0.02$ \\
30 & TrIntraC(d) & $1..4$ & 7.3.8.10 & \texttt{CuPredMode==MODE\_INTRA \&\& (cbf\_cb==$1$; cbf\_cr==$1$)} & $\surd$ & - & $\surd$ \\%& $0.02$ \\
31 & TrInterY(d) & $1..4$ & 7.3.8.10 & \texttt{CuPredMode != MODE\_INTRA \&\& cbf\_luma == $1$} & $\surd$ & - & $\surd$\\% & $0.02$ \\
32 & TrInterC(d) & $1..4$ & 7.3.8.10 & \texttt{CuPredMode!=MODE\_INTRA \&\& (cbf\_cb==$1$; cbf\_cr==$1$)} & $\surd$ & - & $\surd$ \\%& $0.02$ \\
33 & Tr(d) & $1..4$ & 7.3.8.10 & \texttt{cbf\_luma==$1$; cbf\_cb==$1$; cbf\_cr==$1$} & - & $\surd$ & - \\%&  $0.02$ \\
34 & TSF & 4 & 7.4.9.11 & \texttt{transform\_skip\_flag == $1$} & $\surd$ & - & $\surd$\\%& $0.02$ \\
\hline
35 & Bs0  & all& 8.7.2.4 & \texttt{bS == $0$} & $\surd$ & - & $\surd$ \\%& $0.02$ \\
36 & Bs1 & all & 8.7.2.4 & \texttt{bS == $1$} & $\surd$ & - & $\surd$ \\%& $0.02$ \\
37 & Bs2 & all & 8.7.2.4 & \texttt{bS == $2$} & $\surd$ & - & $\surd$ \\%& $0.02$ \\
38 & Bs & all & 8.7.2.4 & always & $\surd$ & $\surd$ & - \\%& $0.02$ \\
%39 & DBFfilt & all & 8.7.2.5.7 & always & $\surd$ & - & $\surd$ \\%& $0.02$ \\
39 & SAO\_Y\_BO & 0 & 7.4.9.3 & \texttt{SaoTypeIdx[Y] == $1$} & $\surd$ & - & $\surd$\\% & $0.02$ \\
40 & SAO\_Y\_EO & 0 & 7.4.9.3 & \texttt{SaoTypeIdx[Y] == $2$} & $\surd$ & - & $\surd$\\% & $0.02$ \\
41 & SAO\_Y & 0 & 7.4.9.3 & \texttt{SaoTypeIdx[Y] != $0$} & $\surd$ & $\surd$ & - \\% & $0.02$ \\
42 & SAO\_C\_BO & 0 & 7.4.9.3 & \texttt{SaoTypeIdx[Cb] == $1$; SaoTypeIdx[Cr] == $1$} & $\surd$ & - & $\surd$ \\%& $0.02$ \\
43 & SAO\_C\_EO & 0 & 7.4.9.3 & \texttt{SaoTypeIdx[Cb] == $2$; SaoTypeIdx[Cr] == $2$} & $\surd$ & - & $\surd$ \\%& $0.02$ \\
44 & SAO\_C & 0 & 7.4.9.3 & \texttt{SaoTypeIdx[Cb] != $0$; SaoTypeIdx[Cr] != $0$} & $\surd$ & $\surd$ & - \\%& $0.02$ \\
45 & SAO\_allComps & 0 & 7.4.9.3 & \texttt{SaoTypeIdx[Y] \&\& SaoTypeIdx[Cb] \&\& SaoTypeIdx[Cr]} & - & - & $\surd$ \\%& $-0.02$ \\
\hline
\end{tabular}
%\end{center}

\end{table*}

\newpage
~
\newpage
~

\newpage 
\bibliographystyle{IEEEtran}

\end{document}